\documentclass[twocolumn,showpacs,preprintnumbers,amsmath,amssymb,nofootinbib]{revtex4}
\usepackage[english]{babel}
\usepackage{graphicx}% Include figure files
\usepackage{dcolumn}% Align table columns on decimal point
\usepackage{bm}% bold math
\usepackage{verbatim}

\newcommand{\ket}[1]{\left| {#1} \right\rangle}
\newcommand{\bra}[1]{\left\langle {#1} \right|}
	
\newcommand{\braket}[2]{\left\langle {#1}\left|{#2}\right.\right\rangle}
\newcommand{\proj}[2]{\left| {#1} \right\rangle\!\left\langle {#2} \right|}

\newcommand{\biket}[2]{\left| {#1} \right\rangle_{I}\left| {#2} \right\rangle_{IV}}

\newcommand{\tr}{\operatorname{Tr}}

\def\slashchar#1{\setbox0=\hbox{$#1$} % set a box for #1
\dimen0=\wd0 % and get its size
\setbox1=\hbox{/} \dimen1=\wd1 % get size of /
\ifdim\dimen0>\dimen1 % #1 is bigger
\rlap{\hbox to \dimen0{\hfil/\hfil}} % so center / in box
#1 % and print #1
\else % / is bigger
\rlap{\hbox to \dimen1{\hfil$#1$\hfil}} % so center #1
/ % and print /
\fi}

\begin{document}

%\nofiles
%\preprint{vs.$8.15$}

\title{Fermionic entanglement that survives a black hole}% Force line breaks with \\
\author{Eduardo Mart\'{i}n-Mart\'{i}nez}%
 \email{martin@imaff.cfmac.csic.es}
\author{Juan Le\'on}
\email{leon@imaff.cfmac.csic.es}
 \homepage{http://www.imaff.csic.es/pcc/QUINFOG/}
\affiliation{%
Instituto de F\'{i}sica Fundamental, CSIC\\
Serrano 113-B, 28006 Madrid, Spain.\\
}

%\affiliation{

\date{\today}% It is always \today, today,

\begin{abstract}
We introduce an arbitrary number of accessible modes when analyzing bipartite entanglement degradation due to Unruh effect between two partners Alice and Rob.  Under the single mode  approximation (SMA) a fermion field  only had a few accessible levels due to Pauli exclusion principle, conversely to bosonic fields which had an infinite number of excitable levels. This was argued to justify entanglement survival in the fermionic case in the SMA infinite acceleration limit. Here we relax SMA. Hence, an infinite number of modes are excited as the observer Rob accelerates, even for a fermion field. We will prove that, despite this analogy with the bosonic case, entanglement loss is limited. We will show that this comes from fermionic statistics through the characteristic structure it imposes on the infinite dimensional density matrix for Rob. Surprisingly, the surviving entanglement is independent of the specific maximally entangled state chosen, the kind of fermionic field analyzed, and the number of accessible modes considered. We shall discuss whether this surviving entanglement goes beyond the purely statistical correlations, giving insight concerning the black hole information paradox.
\end{abstract}

\maketitle

%\section{\label{sec:level1}First-level heading:\protect\\ The line
%break was forced \lowercase{via} \textbackslash\textbackslash}

\section{Introduction}

Studying quantum information in non-inertial settings requires using tools coming from general relativity \cite{Alsingtelep,TeraUeda2,ShiYu,Alicefalls,AlsingSchul,SchExpandingspace,Adeschul,KBr,LingHeZ,ManSchullBlack,PanBlackHoles,AlsingMcmhMil,DH,Steeg,Edu2}. In particular, the Unruh effect \cite{DaviesUnr,Unruh,Takagi,Crispino} --which consists in the emergence of noise when an accelerated observer is describing Minkowski vacuum using Rindler coordinates --affects the possible entanglement that an accelerated observer Rob would share with an inertial observer Alice. The first question to be answered by theory is how much entanglement degrades due to Rob's acceleration $a$.

Some partial answers are in the existing literature \cite{Alicefalls,AlsingSchul,Edu2}. All of them share the shortcomings inherent in the single mode approximation (SMA) \cite{AlsingMcmhMil,Alsingtelep}, which consists in considering only one mode of sharp momentum in the analysis of Unruh degradation. For scalar fields, a Minkowskian maximally entangled state becomes separable in the limit $a\rightarrow\infty$, i.e. the Unruh effect completely destroys entanglement. This is a consequence of the excitation of an unbounded number of modes as Rob accelerates. Contrary to this, finite correlations survive the limit $a\rightarrow\infty$ when considering fermion fields. Pauli exclusion principle --which bounds the maximum number of possible excited modes-- has been argued as the rationale for this \cite{AlsingSchul}. However, that argument was only applied under the (somewhat unphysical) SMA.

We will show that an unbounded number of modes become excited by Unruh effect even for fermion fields if we relax SMA, and so, the above argument ceases to be plausible. Here a fundamental question arises; does fermionic statistics protect the entanglement? Or is this a mere artifact emerging from the SMA?. In this paper we shall show that such entanglement survival is fundamentally inherent in the Fermi-Dirac statistics, and that it is independent of the number of modes considered, of the maximally entangled state we start from, and even of the spin of the fermion field studied.

We will proceed step by step for the sake of clarity. In section \ref{s1} we will introduce the Unruh effect and its impact on entanglement when one of the partners is accelerated. In section \ref{s2} we will express the multimode vacuum and one particle state in the coordinates of an accelerated observer and for two different kinds of fermionic fields (A Dirac field and a ``spinless'' fermion field). After that, we will analyze entanglement degradation for two very different kinds of maximally entangled state of a Dirac field: the case of vacuum entangled with one particle state in section \ref{caso1} and the case of a spin and momentum maximally entangled state in section \ref{caso2}. Both cases were considered under SMA in \cite{Edu2}. Here, we get that, even for the radically different final states obtained in each case, after non-trivial computations entanglement degradation ends up being the same for both and, more importantly, it is independent of the number of modes considered. Then, in section \ref{caso3}, we will investigate degradation for a maximally entangled state of a ``spinless fermion'' field (considered under SMA in \cite{AlsingSchul}). Here, the field and occupation numbers allowed are completely different from the previous cases, hence, the final state is, as well, notably different from them. However, the dependence of entanglement on $a$ turns out to be exactly the same as in the spin $1/2$ cases analyzed before. Finally, we will trace back  this behavior to fermionic statistics (which all the cases share). Specifically, fermionic statistics translates into a peculiar structure in the density matrices for Rob, which is responsible for those striking coincidences. 

Summarizing, although it is true that now we have an infinite number of excited fermionic modes, statistics counterbalances that effect and allows entanglement preservation even at the limit $a\rightarrow\infty$, contrarily to the intuition we would get from the bosonic case. The meaning of this remaining entanglement and its relation with information content of black holes is discussed in the conclusions.

\section{Unruh effect and entanglement decoherence}\label{s1}

The Unruh effect appears when we try to describe fields in the frame of a non-inertial observer. When this happens, an accelerated observer of the minkowskian vacuum, would observe  a thermal particle distribution of temperature $T_U=\hslash a/2\pi k_Bc$. 

To understand where this effect comes from we need to start from a Minkowskian frame and consider the Dirac field expansion in terms of the positive (particle) and the negative (antiparticle) energy solutions of Dirac equation notated $\psi^+_{k,s}$ and $\psi^-_{k,s}$ 
\begin{equation}\label{field}\psi=\sum_{s}\int d^3k\, (a_{k,s}\psi^+_{k,s}+b_{k, s}^\dagger\psi^-_{k,s})\end{equation}
Here, the subscript $k$ notates momentum which labels the modes of the same energy and $s=\{\uparrow ,\downarrow\}$ is the spin label that indicates spin-up or spin-down along the quantization axis. $a_{k,s}$ and $b_{k,s}$ are respectively the annihilation operators for particles and antiparticles, and satisfy the usual anticommutation relations.

For each mode of frequency $k$ and spin $s$ the positive and negative energy modes have the form
\begin{equation}\label{eq2}\psi^\pm_{k,s} =\frac{1}{\sqrt{2\pi k_0}}u^\pm_s(\bm k) e^{\pm i(\bm k\cdot\bm x- k_0t)}\end{equation}
where $u^\pm_s(\bm k)$ is a spinor satisfying the normalization relations $\pm \bar u^\pm_s(\bm k)u^\pm_{s'}(\bm k)=(k_0/m)\delta_{ss'},\bar u^{\mp}_s(\bm k)u^\pm_{s'}(\bm k)=0$.

The modes are classified as particle or antiparticle respect to $\partial_t$ (Minkowski Killing vector directed to the future). The Minkowski vacuum state is defined by the tensor product of each frequency mode vacuum
\begin{equation}\label{vacua}\ket0=\bigotimes_{k,k'}\ket{0_k}^+\ket{0_{k'}}^-\end{equation}
such that it is annihilated by $a_{k,s}$ and $b_{k,s}$ for all values of $s$.

An uniformly accelerated observer viewpoint is described by means of the Rindler coordinates \cite{gravitation}. In order to cover the whole Minkowski space-time, two different set of coordinates are necessary. These sets of coordinates define two causally disconnected regions in Rindler space-time. If we consider that the uniform acceleration $a$ lies on the $z$ axis, the new Rindler coordinates $(t,x,y,z)$ as a function of Minkowski coordinates $(\tilde t,\tilde x,\tilde y,\tilde z)$ are
\begin{equation}\label{Rindlcoordreg1}
a\tilde t=e^{az}\sinh(at),\; a\tilde z=e^{az}\cosh(at),\; \tilde x= x,\; \tilde y= y
\end{equation}
for region I, and
\begin{equation}\label{Rindlcoordreg2}
a\tilde t=-e^{az}\sinh(at),\; a\tilde z=-e^{az}\cosh(at),\; \tilde x= x,\; \tilde y= y
\end{equation}
for region IV.
\begin{figure}\label{fig1}
\includegraphics[width=.45\textwidth]{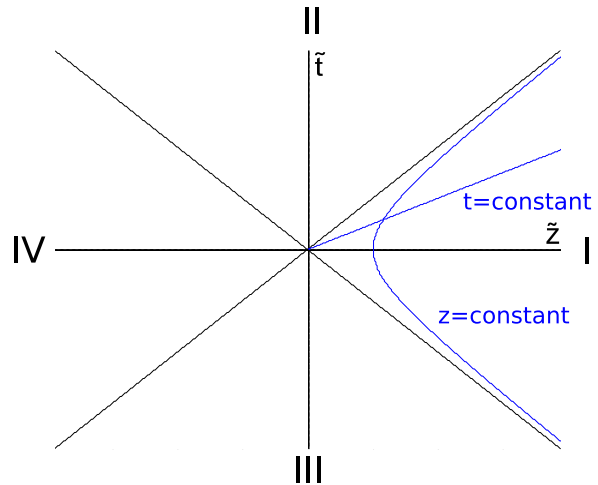}
\caption{Rindler space-time diagram: lines of constant position $z=\text{const.}$ are hyperbolae and all the curves of constant proper time $t$ for the accelerated observer are straight lines that come from the origin. An uniformly accelerated observer Rob travels along a hyperbola constrained to region I}
\end{figure}
As we can see from fig. 1, although we have covered the whole Minkowski space-time with these sets of coordinates, there are two more regions labeled II and III. To map them we would need to switch $\cosh\leftrightarrow\sinh$ in equations \eqref{Rindlcoordreg1},\eqref{Rindlcoordreg2}. In these regions $t$ is a spacelike coordinate and $z$ is a timelike coordinate. However, the solutions of Dirac equation in such regions are not required to discuss entanglement between Alice and an accelerated observer, since he would be constrained to either region I or IV, having no possible access to the opposite regions as they are causally disconnected \cite{Birrell,gravitation,Alicefalls,AlsingSchul}.

The Rindler coordinates $z,t$ go from $-\infty$ to $\infty$ independently in regions I and IV. It means that each region admits a separate quantization procedure with their corresponding positive and negative energy solutions\footnote{Throughout this work we will consider that the spin of each mode is in the acceleration direction and, hence, spin will not undergo Thomas precession due to instant Wigner rotations \cite{AlsingSchul,Jauregui}.} $\{\psi^{I+}_{k,s},\psi^{I-}_{k,s}\}$ and $\{\psi^{IV+}_{k,s},\psi^{IV-}_{k,s}\}$.

Particles and antiparticles will be classified with respect to the future-directed timelike Killing vector in each region. In region I the future-directed Killing vector is
\begin{equation}\label{KillingI}
\partial_t^I=\frac{\partial \tilde t}{\partial t}\partial_{\tilde t}+\frac{\partial\tilde z}{\partial t}\partial_{\tilde z}=a(\tilde z\partial_{\tilde t}+\tilde t\partial_{\tilde z}),
\end{equation}
whereas in region IV the future-directed Killing vector is $\partial_t^{IV}=-\partial_t^{I}$.

This means that solutions in region I, having time dependence $\psi_k^{I+}\sim e^{-ik_0t}$ with $k_0>0$, represent positive energy solutions, whereas solutions in region IV, having time dependence $\psi_k^{I+}\sim e^{-ik_0t}$ with $k_0>0$, are actually negative energy solutions since $\partial^{IV}_t$ points to the opposite direction of $\partial_{\tilde t} $ \cite{AlsingSchul,Birrell}. As I and IV are causally disconnected $\psi^{IV\pm}_{k,s}$ and $\psi^{I\pm}_{k,s}$ only have support in their own regions, vanishing outside them.

Let us denote $(c_{I,k,s},c^{\dagger}_{I,k,s})$ the particle annihilation and creation operators in region I and $(d_{I,k,s},d^{\dagger}_{I,k,s})$ the corresponding antiparticle operators. Analogously we define $(c_{IV,k,s},c^{\dagger}_{IV,k,s}, d_{IV,k,s},d_{IV,k,s}^\dagger)$ the particle/antiparticle operators in region IV.

These operators satisfy the anticommutation relations $\{c_{\text{R},k,s},c^\dagger_{\text{R}',k',s'}\}=\delta_{\text{R}\text{R}'}\delta_{kk'}\delta_{ss'}$ where the subscript R notates the Rindler region of the operator $\text{R}=\{I,IV\}$. All other anticommutators are zero. That includes the anticommutators between operators in different regions of the Rindler space-time.

Taking this into account we can expand the Dirac field in Rindler coordinates analogously to \eqref{field}:
\begin{eqnarray}\label{fieldri}
\nonumber\psi&=&\sum_{s}\int d^3k\, \left(c_{I,k,s}\psi^{I+}_{k,s}+d_{I,k,s}^\dagger\psi^{I-}_{k,s}+c_{IV,k,s}\psi^{IV+}_{k,s}\right.\\*
&&\left.+d_{IV,k,s}^\dagger\psi^{IV-}_{k,s}\right).\end{eqnarray}

Equations \eqref{field} and \eqref{fieldri} represent the expansion of the Dirac field in its modes in Minkowski and Rindler coordinates respectively. We can relate Minkowski and Rindler creation and annihilation operators by taking appropriate inner products and computing the so-called Bogoliubov coefficients \cite{Takagi,Jauregui,Birrell,AlsingSchul}:
\begin{eqnarray}\label{Bogoliubov}
\nonumber a_{k,s}&=&\cos{r}\,c_{I,k,s}-e^{i\phi}\sin r\,d^\dagger_{IV,-k,-s}\\*
b^\dagger_{k,s}&=&\cos{r}\,d^\dagger_{IV,k,s}+e^{-i\phi}\sin r\,c_{I,-k,-s}
\end{eqnarray}
where
\begin{equation}\label{defr}
\tan r=e^{-\pi \frac{k_0c}{a}}
\end{equation}
and $\phi$ is a phase factor that will turn out to be irrelevant for our purposes. Notice that since in Rindler regions I and IV  the temporal Killing vectors pointing to the future have opposite senses, all the magnitudes that are not invariant under time reversal change among regions.

It is shown in the literature \cite{Alicefalls,AlsingSchul,Edu2} that Unruh effect provokes decoherence of entangled states when one of the partners is describing the system from an accelerated frame. As it can be thoroughly seen in \cite{Edu2}, this comes about because the fact of accelerating introduces an horizon in the space-time and, as Rob is always constrained to region I or IV of the Rindler space-time, for a quantum description of Rob's subsystem it will be necessary to trace over the region causally disconnected from Rob. 

It is this partial tracing which transforms the total state (which would be pure for an inertial observer) into a mixed state whose entanglement decreases as Rob acceleration increases. This phenomenon has been called Unruh decoherence in the literature. We are not tackling here the complete problem of expressing the vacuum and the one particle state for the different fields beyond the SMA, because its calculation, although easy, may take a bit long and it is detailedly done in \cite{Edu2}. Instead we will start from the results obtained in \cite{Edu2} for the multimode vacuum expressed as a squeezed state in Rindler coordinates. 

Notice that since the observer Rob is accelerated, his possible measurements are affected by a Doppler-like effect. A discussion of this effect and  how it would affect our study is given in \cite{Edu2}.

\section{Vacuum and 1-Particle states of fermionic fields beyond SMA}\label{s2}

In this section we shall go beyond the single mode approximation to build the vacuum state and the 1-particle excited state for two very different kinds of fermionic fields: First a Dirac field and then a spinless fermion field. Both kinds of fields were analyzed under the SMA in previous literature (the spinless case in \cite{AlsingSchul} and the Dirac field in \cite{Edu2} whose notation we will follow in this paper). 

To begin with, let us consider a discrete number $n$ of different modes of a Dirac field $k_1,\dots,k_n$, labeling with $s_i$ the spin degree of freedom of each mode, so Minkowski multimode vacuum should be expressed as a squeezed state in Rindler coordinates which is an arbitrary superposition of spins and momenta as it is discussed in \cite{Edu2}
\begin{equation}
\label{vacuumCOMP}\ket{0}=\sum_{m=0}^{2n}\sum_{\substack{s_1,\dots,s_{m}\\k_1,\dots,k_{m}}}\!\!\!\!C^{m}_{s_1,\dots,s_{m},k_1,\dots,k_{m}}
\xi_{s_1,\dots,s_{m}}^{k_1,\dots,k_{m}} \biket{\tilde{m}}{\tilde{m}}
\end{equation}
Where, the notation is
\begin{equation}\label{notationmod}
\biket{\tilde i}{\tilde i}\!=\!\biket{s_1,\!k_1;\dots;\!s_i,\!k_i}{-s_1,\!-k_1;\dots;\!-s_i,\!-k_i}
\end{equation}
with
\begin{equation}
\ket{k_1,\!s_1;\dots;\!k_m,\!s_m}_I=c^\dagger_{I,k_m,s_m}\dots c^\dagger_{I,k_1,s_1}\ket{0}_I
\end{equation}
The label outside the kets notates Rindler space-time region, and the symbol $\xi$ is 0 if $\{k_i,s_i\}=\{k_j,s_j\}$ for any $i\neq j$, and it is $1$ otherwise, imposing Pauli exclusion principle constraints on the state (quantum numbers of fermions cannot coincide).

Due to the anticommutation relations of the fermionic operators, terms with different orderings are not independent.
So, without loss of generality, we could choose not to write all the possible orderings in \eqref{vacuumCOMP} selecting one of them instead. In this fashion we will write the elements \eqref{notationmod} with the following ordering criterion:
\begin{eqnarray}\label{ordering}
 \nonumber & k_i\le k_{i+1} &\\
 &k_i=k_{i+1}\Rightarrow s_i=\uparrow,s_{i+1}=\downarrow.&
 \end{eqnarray}
The coefficients $C^m$ are constrained because the Minkowski vacuum should satisfy 
$a_{k_0,s_0}\ket0=0$, $\forall k_0,s_0$. In \cite{Edu2} we showed that imposing this constraint translates into
\begin{equation}\label{coeff2}
C^m=C^0 e^{im\phi}\tan^m r
\end{equation}
Where $\tan r=\exp\left(-\pi k_0c/a\right)$. $C^m$ is independent of $s_i$ and $k_i$. Therefore, we obtain the vacuum state by substituting \eqref{coeff2} in \eqref{vacuumCOMP} and factoring the coefficients out of the $k_i,s_i$ summation.
\begin{equation}\label{vacuumCOMP2}
\ket{0}=\sum_{m=0}^{2n}C^m\!\!\!\sum_{\substack{s_1,\dots,s_m\\k_1,\dots,k_m}}\!\!\!\!\xi_{s_1,\dots,s_m}^{k_1,\dots,k_m} \biket{\tilde m}{\tilde m}\\
\end{equation}
The only parameter not fixed yet is $C^0$. We can fix it imposing the normalization of the Minkowski vacuum in Rindler coordinates $\braket{0}{0}=1$, 
\cite{Alicefalls,AlsingSchul,Edu2} which implies
\begin{equation}\label{normalispin}|C^0|=\cos^{2n}r\end{equation}
Where we have taken into account our ordering choice explained above. This expression is explicitly derived in the appendix \ref{ap1}.

Eq. \eqref{normalispin} gives the value of $C^0$ except for a global phase. Next, the 1-particle state can be worked out translating the Minkowski one particle state $\ket{k,s}=a^\dagger_{k,s}\ket0$ into Rindler coordinates
\begin{equation}\label{onepart2}
\ket{k,s}=\sum_{m=0}^{2n-1}A^m\sum_{\substack{s_1,\dots,s_m\\k_1,\dots,k_m}}\!\!\!\!\xi_{s_1,\dots,s_m,s}^{k_1,\dots,k_m,k} \biket{\tilde m;k,s}{\tilde m}
\end{equation}
Where
\begin{equation}\label{Am}
A^m=(C^m\cos r+C^{m+1}e^{-i\phi}\sin r)
\end{equation}
and the notation $\ket{\tilde m;s,k}_I$, consequently with \eqref{notationmod}, means the ordered version of 
$\ket{s_1,\!k_1;\dots;\!s_n,\!k_n;k,s}_I$. 

Another different kind of field that we are going to consider appears by neglecting spin keeping the fermionic statistics (like considering Grassman scalar 
fields). This kind of field is used under the SMA in the literature \cite{AlsingSchul}. Here we will relax such approximation and will analyze Unruh decoherence when we allow $n$ different momenta $k_i$. Barring spin, the Minkowski multimode vacuum state would be expressed as
\begin{equation}\label{vacuzero}
\ket{0}=\sum_{m=0}^n\sum_{k_1,\dots,k_m}\xi_{k_1,\dots,k_m}\hat C^m_{k_1,\dots,k_m}\ket{\tilde m}_I\ket{\tilde m}_{IV}
\end{equation}
where, in this occasion $\ket{\tilde m}_I\ket{\tilde m}_{IV}=\ket{k_1,\dots,k_m}_I$ $\ket{-k_1,\dots,-k_m}_{IV}$. Using the same procedures as for the spin $1/2$ 
case \eqref{vacuumCOMP} we can prove that all the coefficients are independent of $k_i$ and can be related to $\hat C^0$ as in \eqref{coeff2}, $\hat C^m=\hat C^0 e^{im
\phi}\tan^m r$. We can now fix $\hat C^0$ imposing the normalization relation $\braket00=1$ giving
\begin{equation}\label{normalizeropre}
\hat C^0=\left[\sum_{m=0}^n\chi_m\tan^{2m}r\right]^{-1/2}
\end{equation}
where
\begin{equation}\label{XII}
\chi_m\equiv\sum_{k_1,\dots,k_m}\xi_{k_1,\dots,k_m}=\binom{n}{m}
\end{equation}
Such that \eqref{normalizeropre} can be simplified to
\begin{equation}\label{normalizero}
\hat C^0=\left[\sum_{m=0}^n\binom{n}{m}\tan^{2m}r\right]^{-1/2}=\cos^n r
\end{equation}
 
Finally, the one particle state $a_k^\dagger\ket0$ is
\begin{equation}\label{vacuuno}
\ket{k}=\sum_{m=0}^{n-1}\hat A^m\!\!\sum_{k_1,\dots,k_m}\!\!\xi_{k_1,\dots,k_m,k}\biket{\tilde m,k}{\tilde m}
\end{equation}
where $\hat A^m$ has the expression \eqref{Am} but substituting $C^m$ by $\hat C^m$.

\section{Entanglement degradation for a $\ket{00}+\ket{11}$  entangled state of a Dirac field beyond SMA}\label{caso1}

In the following we will analyze Unruh entanglement degradation in various settings corresponding to different maximally entangled states of fermion fields. First we consider the following state in Minkowskian coordinates
\begin{equation}\label{minkowstate1}\ket\Psi=\frac{1}{\sqrt2}\big(\ket{0}_A\ket{0}_R+\ket{k_A,s_A}_A\ket{k_R,s_R}_R\big)\end{equation}
The density matrix for the accelerated observer Rob is obtained after expressing Rob's state in Rindler coordinates --which means using \eqref{vacuumCOMP} and \eqref{onepart2} in Rob's part of \eqref{minkowstate1}-- and afterwards, tracing over Rindler's region $IV$ since Rob is causally disconnected from it and he is not to extract any information from beyond the horizon. Following this procedure we obtain the density matrix $
\rho=$
\begin{eqnarray}\label{densmat1}
\nonumber&&\!\!\!\!\!\!\!\!\!\!\!\!\!\!\frac{1}{2}\Big[\sum_{m=0}^{2n}\!\Big(D_{0}^m\!\!\!\sum_{\substack{s_1,\dots,s_m\\k_1,\dots,k_m}}\!\!\!
\xi_{s_1,\dots,s_m}^{k_1,\dots,k_m}\ket{0}_A\ket{\tilde m}_I\bra{0}_A\bra{\tilde m}_I\!\!\Big)\\*
\nonumber&&\!\!\!\!\!\!\!\!\!\!\!\!\!\!+\sum_{m=0}^{2n-1}\Big(D_{1}^m\!\!\!\sum_{\substack{s_1,\dots,s_m\\k_1,\dots,k_m}}\!\xi_{s_1,\dots,s_m,s_R}^{k_1,\dots,k_m,k_R}
\ket{0}_A\ket{\tilde m}_I\bra{k_A,s_A}_A\\*
\nonumber&&\!\!\!\!\!\!\!\!\!\!\!\!\!\!\times\!\bra{\tilde m;k_R,s_R}_I\!\!\Big)\!+\!\!\sum_{m=0}^{2n-1}\!\!\Big(D_{2}^m\!\!\!\!\sum_{\substack{s_1,\dots,s_m\\k_1,\dots,k_m}}\!\!\!\!
\xi_{s_1,\dots,s_m,s_R}^{k_1,\dots,k_m,k_R}\ket{k_A,s_A}_A\\*
&&\!\!\!\!\!\!\!\!\!\!\!\!\!\!\times\ket{\tilde m;k_R,s_R}_I\bra{k_A,s_A}_A\bra{\tilde m;k_R,s_R}_I\Big)\Big]+(\text{H.c.})_{_{\substack{\text{non-}\\\text{diag.}}}}
\end{eqnarray}
Where $(\text{H.c.})_{\text{non-diag.}}$ means Hermitian conjugate of only the non-diagonal terms and
\begin{equation}\label{Des}
D_i^m=|C^0|^2\frac{\tan^{2m}r}{\cos^i r}
\end{equation}
with $i=0,1,2$. The derivation of \eqref{densmat1} can be found in the appendix \ref{ap2}. 

Notice that as Rob accelerates, the state becomes mixed with all the available modes $(k_1,\dots,k_n)$ excited. This contrasts with the Minkowskian state 
\eqref{minkowstate1} where only one Rob mode is excited $(k_R,s_R)$. Notice also that the Hilbert space dimension has changed from $2\times 2$ to 
$2n\times2n$.

We will compute the negativity as a function of $a$ as a measure of the state entanglement, the negativity is the sum of all the negative eigenvalues of the partial transpose of $\rho$.

The partial transpose of \eqref{densmat1} has a $2\times2$ and $1\times1$ blocks structure. Each eigenvalue in the $1\times1$ blocks is non-negative (since 
$D_i^m\ge0$), so we are interested in the $2\times2$ which are the ones that may have negative eigenvalues. These $2\times2$ blocks expressed in the basis
\begin{equation}\label{blocksbasis}
\Big\{\ket0_{A}\ket{\tilde m;k_R,s_R}_I,\ket{s_A,k_A}_{A}\ket{\tilde m}_I\Big\}_{m=0}^{2n-1}
\end{equation}
are of the form
\begin{equation}\label{blocks}
\frac12
\left(\begin{array}{cc}
D^{m+1}_0 & \pm D_1^m\\
\pm D_1^m & 0
\end{array}\right)
\end{equation}
There is no $D_2^m$ element because it goes with $\ket{k_A,s_A}_A\ket{\tilde m;k_R,s_R}_I\bra{k_A,s_A}_A\bra{\tilde m;k_R,s_R}_I$ 
which cannot have any element within our block as Pauli exclusion principle imposes $k_R,s_R\not\in\left\{k_i,s_i\right\}_{i=1,\dots,m}$.

Each $2\times2$ block of \eqref{blocks} appears a number of times $B_m$ given by
\begin{equation}\label{Nblocks}B_m=\binom{2n-1}{m}.
\end{equation}
The derivation of this formula can be found in the appendix \ref{ap1}.

Using \eqref{Des}, the negative eigenvalue of each block can be expressed
\begin{equation}\label{neigenb}
|\lambda^-_m|=\frac12|C^0|^2\,\tan^{2m} r
\end{equation}
where $C_0$ is given by \eqref{normalispin}. Therefore, the negativity is expressed as the summation of the negative eigenvalue of each block $|\lambda_m^-|$ multiplied by the number of times $B_m$ that that block appears in the partially transposed density matrix. The 
summation of the series is
\begin{equation}\label{negativitypre1}
\mathcal{N}=\sum_{m=0}^{2n-1} B_m|\lambda^-_m|=\frac{\cos^{4n}r}{2}\sum_{m=0}^{2n-1}\binom{2n-1}{m}\tan^{2m}r 
\end{equation}
but this result can be easily simplified to
\begin{equation}\label{negativitypre}
\mathcal{N}=\frac12\cos^2 r
\end{equation}
which is independent of the number of modes $n$ that we have considered. This surprising result shows that, even if we consider more than one mode in our 
field analysis, the entanglement degradation due to Unruh effect is the same as considering only one mode as it is done in \cite{Edu2}. In other words, despite the fact that all the available modes are excited when Rob accelerates \eqref{densmat1}, the quantum correlations behave as if we were considering only one possible mode for the field. This is a consequence of the peculiar structure of the density matrix for Rob, being the fermionic nature of the field the final responsible of this structure \eqref{blocksbasis}.

\section{Entanglement degradation for a spin and momentum entangled state of a Dirac field beyond SMA}\label{caso2}

If instead of \eqref{minkowstate1} we start from a Bell momentum-spin state in Minkowski coordinates
\begin{equation}\label{minkowstate2}\ket\Psi=\frac{1}{\sqrt2}\Big(\ket{k_A^1,s_A^1}_A\ket{k^1_R,s^1_R}_R+\ket{k^2_A,s^2_A}_A\ket{k^2_R,s^2_R}_R\Big)
\end{equation}
As it can be seen in the appendix \ref{ap2} the density matrix for Rob takes the form $\rho=$
\begin{eqnarray}\label{densmat2}
&&\!\!\!\!\!\!\!\!\!\nonumber\sum_{m=0}^{2n-1}\!\!\frac{D_{2}^m}{2}\!\!\!\!\!\sum_{\substack{s_1,\dots,s_m\\k_1,\dots,k_m}}\!\!\!\!
\Big(\xi_{s_1,\dots,s_m,s_R^1}^{k_1,\dots,k_m,k_R^1}\!\ket{k^1_A,s^1_A}_{\!A}\!\ket{\tilde m;k^1_R,s^1_R}_I\bra{k^1_A,s^1_A}_A\\*
&&\!\!\!\!\!\!\!\!\nonumber\times\!\bra{\tilde m,k^1_R,s^1_R}_I+\xi_{s_1,\dots,s_m,s_R^2}^{k_1,\dots,k_m,k_R^2}\ket{k^2_A,s^2_A}_A\ket{\tilde m;k^2_R,s^2_R}\\*
&&\!\!\!\!\!\!\!\!\nonumber\times\bra{k_A^2,s_A^2}_A\bra{\tilde m;k^2_R,s_R^2}_I+
\xi_{s_1,\dots,s_m,s_R^1,s_R^2}^{k_1,\dots,k_m,k_R^1,k_R^2}\ket{k^1_A,s^1_A}_A\\*
&&\!\!\!\!\!\!\!\!\times\ket{\tilde m;k_R^1,s_R^1}_I\bra{k^2_A,s^2_A}_A\bra{\tilde m;k^2_R,s^2_R}_I\Big)+(\text{H.c.})_{_{\substack{\text{non-}\\\text{diag.}}}}
\end{eqnarray}
Analogously to \eqref{densmat1}, the partial transpose of \eqref{densmat2} has a $2\times2$ and $1\times1$ blocks structure. Again, we are interested in the 
$2\times2$ blocks --the ones that may have negative eigenvalues.-- These blocks expressed in the basis
\begin{equation}\label{blocksbasis2}
\Big\{\ket{k^1_A,s^1_A}_{A}\ket{\tilde m;k^2_R,s^2_R}_I,\ket{s^2_A,k^2_A}_{A}\ket{\tilde m,k^1_R,s^1_R}_I\Big\}_{m=0}^{2n-2}
\end{equation}
are of the form
\begin{equation}\label{blocks2}
\frac12
\left(\begin{array}{cc}
0 & \pm D_2^m\\
\pm D_2^m & 0
\end{array}\right)
\end{equation}
Notice that there is no diagonal elements in the block because the terms that would go in the diagonal are forbidden by Pauli exclusion principle,  which imposes 
that $k^1_R,s^1_R;k^2_R,s^2_R\not\in\left\{k_i,s_i\right\}_{i=1,\dots,m}$. This time, each $2\times2$ block of the form \eqref{blocks2} appears a number of 
times $B'_m$ given by
\begin{equation}\label{Nblocks2}B'_m=\binom{2n-2}{m}
\end{equation}
(See appendix \ref{ap1}). The negative eigenvalue of each block is
\begin{equation}\label{neigen2}
|\lambda^-_m|=\frac{D_2^m}{2}=\frac{\cos^{4n-2}r}{2}\tan^{2m}r
\end{equation}
where $C_0$ has been substituted by \eqref{normalispin}. Therefore, the negativity results
\begin{equation}\label{negativity2pre}
\mathcal{N}=\sum_{m=0}^{2n-2} B'_m|\lambda^-_m| =\frac{\cos^{4n-2}r}{2}\sum_{m=0}^{2n-2}\binom{2n-2}{m}  \tan^{2m}r
\end{equation}
This can be readily simplified to
\begin{equation}\label{negativity2pre}
\mathcal{N}= \frac12\cos^2 r
\end{equation}
Strikingly we run into the same simple result as above \eqref{negativitypre}. Even starting from a spin Bell state, the entanglement is degraded by Unruh effect in the same way as in the previous case.

\section{Entanglement degradation for  $\ket{00}+\ket{11}$  entangled state of a spinless fermion field beyond SMA}\label{caso3}

Now we can go one step further neglecting spin and consider a spinless field on which we have imposed the fermionic statistics. The maximally entangled state of the vacuum and one particle in this setting
\begin{equation}\label{minkowstate3}
\ket\Psi=\frac{1}{\sqrt{2}}\Big(\ket{0}_A\ket{0}_R+\ket{k_A}_A\ket{k_R}_R\Big)
\end{equation}
As it is discussed in appendix \ref{ap2}, this leads to the following density matrix for the accelerated observer Rob after using expressions \eqref{vacuzero} and \eqref{vacuuno} and after tracing over Rindler's region $IV$ $\rho=$
\begin{eqnarray}\label{densmat3}
\nonumber&&\!\!\!\!\!\!\!\frac{1}{2}\Big[\sum_{m=0}^{n}\!\hat D_{0}^m\!\!\!\sum_{k_1,\dots,k_m}\!\!\!\xi_{k_1,\dots,k_m}\ket{0}_A\ket{\tilde m}_I\bra{0}_A\bra{\tilde m}_I+
\sum_{m=0}^{n-1}\Big(\hat D_{1}^m\\*
\nonumber&&\!\!\!\!\!\!\!\!\!\sum_{k_1,\dots,k_m}\!\xi_{k_1,\dots,k_m,k_R}\ket{0}_A\ket{\tilde m}_I\bra{k_A}_A\bra{\tilde m;k_R}_I+\hat D_{2}^m\sum_{k_1,\dots,k_m}\\*
\nonumber&&\!\!\!\!\!\!\!\xi_{k_1,\dots,k_m,k_R}\ket{k_A}_A\ket{\tilde m;k_R}_I\bra{k_A}_A\bra{\tilde m;k_R}_I\Big)\Big]+(\text{H.c.})_{_{\substack{\text{non-}\\\text{diag.}}}}\\*
\end{eqnarray}
where $\hat D^,_i$ is given by the expression \eqref{Des} but substituting $C^0$ by $\hat C_0$. 

Analogously to \eqref{densmat1} and \eqref{densmat2}, The partial transpose of \eqref{densmat3} has a $2\times2$ and $1\times1$ blocks structure. The 
$2\times2$ blocks expressed in the basis
\begin{equation}\label{blocksbasis3}
\Big\{\ket{0}_{A}\ket{\tilde m;k_R}_I,\ket{k_A}_{A}\ket{\tilde m}_I\Big\}_{m=0}^{n-1}
\end{equation}
would have the form 
\begin{equation}\label{blockszero}
\frac12
\left(\begin{array}{cc}
\hat D^{m+1}_0 & \pm \hat D_1^m\\
\pm \hat D_1^m & 0
\end{array}\right)
\end{equation}
The main difference with \eqref{blocks} is that $\hat C_0$ a different value \eqref{normalizero} instead of $C^0$ given by \eqref{normalispin}. Here, $\hat D_2^m$ does not appear because Pauli exclusion principle imposes that $k_R\not\in\left\{k_i\right\}_{i=1,\dots,m}$. Now, each $2\times2$ block multiplicity is
\begin{equation}\label{Nblocks3}B''_m=\binom{n-1}{m}
\end{equation}
(See appendix).
The negative eigenvalue of each block is given by the same expression \eqref{neigenb} but $C_0$ is now given by 
\eqref{normalizero}, which is to say
\begin{equation}\label{neigenb0}
|\lambda^-_m|=\frac12|\hat C^0|^2\,\tan^{2n} r=\frac12\cos^{2m}\,\tan^{2m} r
\end{equation}
 We can compute the negativity
\begin{equation}\label{negazeropre}
\mathcal{N}=\sum_{m=0}^{n-1} B''_m|\lambda^-_m|=\frac12\cos^{2n} r\sum_{m=0}^{n-1}\binom{n-1}{m}\tan^{2m}r
\end{equation}
At this point, the reader might not be surprised by the resulting negativity after straightforward simplification
\begin{equation}\label{negazero}
\mathcal{N}=\frac{1}{2}\cos^2 r
\end{equation}
which is the same result as in the cases \eqref{minkowstate1} and 
\eqref{minkowstate2}. Again, entanglement degradation due to Unruh effect is 
the same as considering one mode of a Dirac field \cite{Edu2}.

\section{Conclusions and comments}
Let us summarize our results so far. We have studied entanglement degradation by Unruh effect as Rob accelerates beyond the single mode approximation and  three different Minkowskian maximally entangled states: 1) Vacuum-vacuum plus one-particle-one-particle maximally entangled state of a Dirac field, 2) spin-momentum Bell state for a Dirac field, 3) Vacuum-vacuum plus one-particle-one-particle maximally entangled state of a spinless fermion field. In spite of the essential differences among these states, the negativity degrades in exactly the same way for any acceleration. This result may look surprising considering that this is the same degradation obtained under the single mode approximation  \cite{AlsingSchul, Edu2} but as it is discussed in this paper, this is an outcome of fermionic statistics.

In the bosonic case acceleration excites an infinite number of modes, and this completely degrades the entanglement in the limit $a\rightarrow\infty$. Although one could expect the same behavior here --as an infinite number of modes is also excited when we let $n\rightarrow\infty$-- our results show that some entanglement is preserved for $a\rightarrow\infty$. It is remarkable that the entanglement degradation coincides for all the different cases considered, with independence of the number of modes $n$.

This striking result can be traced back to the fanciful block structure of Rob density matrix, which produces the same negativity even when the characteristics of the entangled states (and even the field) change. The culprit of this structure is fermionic statistics, (as we have discussed after \eqref{blocks}, \eqref{blocks2}, \eqref{blockszero}) which is responsible for the identical, and somewhat unforeseen, negativity behavior. This is a global feature of maximally entangled states of fermion fields and not a consequence of the specific cases chosen and the number of modes considered. 

So, $\mathcal{N}\rightarrow 
1/4$ when $a\rightarrow\infty$, and this happens independently of the number of modes of the field that we are considering, of the starting maximally entangled state, and even of the spin of the field which we study. What all the cases have in common is the fermionic statistics itself, so, widening the margin for Unruh degradation for fermionic fields will not affect entanglement degradation.

Notice that a very different scenario would come from a setting in which we erase partial information for the state as Rob accelerates (e.g. angular momentum). In that case, it was shown that entanglement degradation is greater than in the cases where all the information is taken into account \cite{Edu2}, but this has more to do with this erasure of information than which the fermionic nature of the states.

One question immediately arises from these results; Are  the remaining correlations purely statistical? In other words, does any useful information really survive the limit $a\rightarrow\infty$?. As all the states undergo the same degradation, everything points that statistics is the only information which survives this limit. 

Furthermore, the limit $a\rightarrow\infty$ can be understood as considering an observer moving in a trajectory arbitrarily close to the event horizon of a Schwarzschild black hole \cite{Alicefalls}. So, even if Alice is free falling into a black hole and Rob stands at the event horizon, a fixed degree of entanglement survives  to Unruh decoherence. Apart from the interest of describing the entanglement between accelerated observers, the regularity and universality of our result $(N=(1/2) \cos^2 r)$ could be a useful hint in the solution of the information paradox in black holes and deserves further investigation in future works.

\section{Acknowledgements}

This work was partially supported by the Spanish MICINN Project FIS2008-05705/FIS. E. M-M was partially supported by the CSIC JAE-PREDOC2007 Grant.

\appendix
\section{Derivation of $C_0$ and the combinatory formulae}\label{ap1}

To derive $C_0$ except for a global phase, we impose the normalization of the vacuum state in Rindler coordinates $\braket00=1$, from \eqref{vacuumCOMP2}, we see that this means that
\begin{equation}\label{normalispinap}C^0=\left[\sum_{m=0}^n\Upsilon_m\tan^{2m}r+\sum_{m=n+1}^{2n}\Upsilon_{2n-m}\tan^{2m}r\right]^{-1/2}\end{equation}
where
\begin{equation}\label{upsilon}\Upsilon_m=\sum_{\substack{s_1,\dots,s_m\\k_1,\dots,k_m}}\!\!\!\!\xi_{s_1,\dots,s_m}^{k_1,\dots,k_m}
\end{equation}
Now, we are going to show that \eqref{upsilon}, has the form
\begin{equation}\label{upsilonap}\Upsilon_m=\sum_{p=0}^{\lfloor \frac{m}
{2}\rfloor}\binom{n-p}{m-2p}\binom{n}{p}2^{m-2p}
\end{equation}
To see how this expression comes from Pauli exclusion principle, we have to read $p$ as an index that represents the number of possible spin pairs ($k_i=k_{i+1},s_i=\uparrow,s_{i+1}=\downarrow$) which can be formed, and goes from $0$ to the integer part of $m/2$, and then
\begin{itemize}
\item The combinatory number $\binom{n-p}{m-2p}$ represents the possible combinations of modes that can be formed taking into account that $p$ different momenta $k_i$ are not available since they are already occupied by the $p$ pairs. Hence, it is given by the combinations of the $n-p$ available momenta taken $m-2p$ at time, since $m-2p$ is the number of free momentum `slots' (the total number of different momenta $m$ minus the number of positions taken by pairs $2p$).
\item The combinatory factor $\binom{n}{p}$ represents the different possible combinations for the configuration of the $p$ pairs, which have $n$ possible different momenta to be combined among them without repetition and in a particular order.
\item The factor $2^{m-2p}$ represents the possible combination for the spin degree of freedom of each mode. As a spin pair only admits one spin configuration, only the unpaired modes will give different spin contributions, so the factor is $(2S+1)^{m-2p}$ giving the formula \eqref{upsilon}
\end{itemize}

After some lengthy but elementary algebra we can see that
\begin{equation}\label{upsilon22}\Upsilon_m=\binom{2n}{m}
\end{equation}
and using the property $\binom{a}{a-b}=\binom{a}{b}$, 
we can express \eqref{normalispinap} as
\begin{equation}
C^0=\left[\sum_{m=0}^{2n}\binom{2n}{m}\tan^{2m}r\right]^{-1/2}=\cos^{2n}r
\end{equation}
which is equation \eqref{normalispin}.

Now we will do the same for the equation \eqref{Nblocks}. This formula takes into account the number of two by two blocks of the form \eqref{blocks}. Taking a look at the basis in which those blocks are expressed \eqref{blocksbasis}, we can see that the expression for $B_m$ is given by two terms:
\begin{itemize}
\item  The number of possible combinations of $m$ modes with $n$ possible different momenta $k_i$ and two possible spins $s_i$ according to Pauli exclusion principle as in \eqref{upsilonap}.
\item A negative contribution which comes from excluding those combinations in which $\{k_R,s_R\}$ coincides with any $\{k_i,s_i\}$, which means excluding the number of combinations in \eqref{upsilonap} which have one of their values fixed to $\{k_i,s_i\}=\{k_R,s_R\}$. This number is given by the combinatory number $\binom{2n-1}{m-1}$ provided that $m>0$ and it is zero if $m=0$.
\end{itemize}

To see where this negative contribution comes from let us assume that it is $\{k_i,s_i\}$ the mode which coincides with $\{k_R,s_R\}$ we will have $2n-1$ possible choices for each $\{k_{j\neq i},s_{j\neq i}\}$ ($2$ values for $s$ and $n$ for $k$ excepting $k_i,s_i$ due to Pauli exclusion principle). This happens for all the combinations of all the possible values $\{k_j,s_j\}$ with $j\neq i$. Hence, as there are $m$ modes and one of them is fixed $k_i=k_R, s_i=s_R$, we have to consider the combinations of $2n-1$ elements taken $m-1$ at time.

If $m>n$ the situation is equivalent to having $m'=2n-m$. Since having more modes $m$ than possible values of $k_i$ we are forced to have $n-m$ pairs and we lose freedom to combine the available modes.

Now if we compute
\begin{equation}\label{Nblocksap}B_m=\Upsilon_m-\binom{2n-1}{m-1}=\binom{2n}{m}-\binom{2n-1}{m-1}
\end{equation}
After some basic algebra we obtain
\begin{equation}\label{Nblocksap}B_m=\binom{2n-1}{m}\end{equation}
which is expression \eqref{Nblocks}

The derivation of expression \eqref{Nblocks2} is quite straightforward considering the one above. Looking at the basis of the $2\times2$ blocks \eqref{blocksbasis2} we can see that this case would be exactly the same as the previous one but now $\{k_i,s_i\}$ cannot coincide neither with $\{k_R^1,s_R^1\}$ nor $\{k_R^2,s_R^2\}$. Repeating the same reasoning as before we have to do three operations as follows
\begin{itemize}
\item Discounting the combinations which have a coincidence $\{k_i,s_i\}=\{k_R^1,s_R^1\}$ from the total number \eqref{upsilonap} and obtain the expression \eqref{Nblocksap}
\item Subtracting the combinations with coincidences $\{k_j,s_j\}=\{k_R^2,s_R^2\}$
\item Taking into account that we have subtracted twice the cases in which we have double coincidences, we need to add the number of double coincidences once to compensate it.
\end{itemize}
 The number of cases with double coincidences (which require $m>1$) is the combinatory number $\binom{2n-2}{m-2}$, as we have $2n$ possible spins and momenta minus the two fixed possibilities ($\{k_i,s_i\}=\{k_R^1,s_R^1\}$ and $\{k_j,s_j\}=\{k_R^2,s_R^2\}$) and $m$ modes being 2 of them fixed. Taking this into account
\begin{equation}\label{apend2}
B'_m=\Upsilon_m -\binom{2n-1}{m-1}-\binom{2n-1}{m-1} + \binom{2n-2}{m-2}
\end{equation}
This expression can be simplified to
\begin{equation}\label{Nblocks2ap}B'_m=B_m-\binom{2n-2}{m-1}=\binom{2n-2}{m}
\end{equation}
which is \eqref{Nblocks2}

For the spinless fermion field, equation \eqref{XII}, which have the form
\begin{equation}\label{XIIap}
\chi_m\equiv\sum_{k_1,\dots,k_m}\xi_{k_1,\dots,k_m}=\binom{n}{m}
\end{equation}
corresponding to the possible combinations of m values of $k_i$ imposing that $k_i\neq k_j$ if $j\neq i$ (which is the translation of Pauli exclusion principle to spinless modes). This can be readily obtained taking into account that the $n$ possible values of $k_i$ should be combined without repetition in a particular ordering of the $m$ modes, so the possible combinations are simply the combinatory number $\binom{n}{m}$

Equation \eqref{Nblocks3} can be easily obtained taking into account that the number of $2\times2$ blocks \eqref{blocksbasis3} is given by the number of mode combinations allowed by Pauli principle \eqref{XIIap}, subtracting the terms having $k_i=k_R$. The number of possible $k_j$ values allowed for the rest $m-1$ modes having fixed $k_i=k_R$ is $n-1$, so the number of combinations we must subtract is the combinatory number $\binom{n-1}{m-1}$, obtaining
 \begin{equation}\label{Nblocks3ap}B''_m=\binom{n}{m}-\binom{n-1}{m-1}=\binom{n-1}{m}
\end{equation}
which is \eqref{Nblocks3}

\section{Density matrix construction}\label{ap2}

In this appendix we will derive expressions \eqref{densmat1}, \eqref{densmat2}, \eqref{densmat3} for the density matrix of the system Alice-Rob.

Using expression \eqref{vacuumCOMP2} we see that the Alice-Rob Minkowskian operator $P_{00}\equiv\proj{0;0}{0;0}$ when Rob is accelerating translates into
\begin{eqnarray}\label{0000}
\nonumber P_{00}\!\!&=&\!\!\sum_{m=0}^{2n}\sum_{l=0}^{2n}C^m (C^{l})^*\!\!\!\sum_{\substack{s_1,\dots,s_m\\k_1,\dots,k_m}}\!\!\!\!\xi_{s_1,\dots,s_m}^{k_1,\dots,k_m}\sum_{\substack{s'_1,\dots,s'_{l}\\k_1,\dots,k'_{l}}}\!\!\!\!\xi_{s'_1,\dots,s'_l}^{k'_1,\dots,k'_l}\\*
\!\!&& \!\!\times\ket{0}_A\biket{\tilde m}{\tilde m}\langle{\tilde l}|_{IV}\langle{\tilde l}|_{I}\bra{0}_A
\end{eqnarray}
where
\begin{eqnarray}
\nonumber\langle{\tilde l}|_{I}&=&\bra{k'_1,s'_1;\dots; k'_l,s'_l}_I=\bra0_I c_{I,k'_1,s'_1}\dots c_{I,k'_m,s'_m}\\*
\nonumber\langle{\tilde l}|_{IV}&=&\bra{-k'_1,-s'_1;\dots; -k'_l,-s'_l}_I\\*
&=&\bra0_{IV} c_{IV,-k'_1,-s'_1}\dots c_{IV,-k'_m,-s'_m}
\end{eqnarray}
In expression \eqref{0000}, and below in \eqref{1111},  bras and kets refer to Alice's mode in Minkowski coordinates and Rob's mode in Rindler coordinates.

Now, using expression \eqref{onepart2} we can write the operator $P_{11}^{ij}\equiv | k^i_A,s^i_A;k^i_R,s^i_R\rangle\langle k^j_A,s^j_A;k^j_R,s^j_R|$
\begin{eqnarray}\label{1111}
\nonumber P_{11}^{ij}\!\!&=&\!\!\sum_{m=0}^{2n-1}\sum_{l=0}^{2n-1}A^m(A^l)^*\sum_{\substack{s_1,\dots,s_m\\k_1,\dots,k_m}}\!\!\!\!\xi_{s_1,\dots,s_m,s^i_R}^{k_1,\dots,k_m,k^i_R}\\*
\nonumber&&\times\sum_{\substack{s'_1,\dots,s'_{l}\\k_1,\dots,k'_{l}}}\!\!\!\!\xi_{s'_1,\dots,s'_l,s^j_R}^{k'_1,\dots,k'_l,k^j_R}
 \ket{k^i_A,s^i_A}_A\biket{\tilde m;k^i_R,s^i_R}{\tilde m}\\*
&&\times\,\langle \tilde l|_{IV}\langle \tilde l;k^j_R,s^j_R|_I\langle{k^j_A,s^j_A}|_A
\end{eqnarray}
where $A^m$ is given by \eqref{Am}.

Notice that the objects $\ket{\tilde m;k^i_R,s^i_R}_I$ represent the appropriate ordering of the elements inside with its sign, taking the criterion \eqref{ordering} into account.

Now we can use expressions \eqref{vacuumCOMP2} and \eqref{onepart2} to obtain the operator $P_{01}\equiv\proj{00}{k_A,s_A;k_R,s_R}$ as it is expressed when Rob is describing the world using Rindler coordinates.
\begin{eqnarray}\label{0011}
\nonumber P_{01}\!\!&=&\!\!\sum_{m=0}^{2n}\sum_{l=0}^{2n-1}C^m (A^{l})^*\!\!\!\sum_{\substack{s_1,\dots,s_m\\k_1,\dots,k_m}}\!\!\!\!\xi_{s_1,\dots,s_m}^{k_1,\dots,k_m}\sum_{\substack{s'_1,\dots,s'_{l}\\k_1,\dots,k'_{l}}}\!\!\!\!\xi_{s'_1,\dots,s'_l,s_R}^{k'_1,\dots,k'_l,k_R}\\*
\!\!&& \!\!\times\ket{0}_A\biket{\tilde m}{\tilde m}\langle{\tilde l}|_{IV}\langle{\tilde l;k_R,s_R}|_{I}\bra{0}_A
\end{eqnarray}

After obtaining the expressions for the operators $P_{00},P_{11},P_{01}$ we can write the density matrix associated with the state \eqref{minkowstate1} in Rindler coordinates for Rob, 
\begin{equation}\label{roimp1}
\rho=\frac12\left(P_{00}+P_{01}+P_{01}^\dagger+P^{ii}_{11}\right)
\end{equation}
Where for $P_{11}^{ii}$ we are considering $\{k^i_R,s^i_R\}=\{k^j_R,s^j_R\}\equiv \{k_R,s_R\}$ and $\{k^i_A,s^i_A\}=\{k^j_A,s^j_A\}\equiv \{k_A,s_A\}$.

We can do the same to obtain the density matrix associated with \eqref{minkowstate2} in Rindler coordinates for Rob
\begin{equation}\label{roimp2}
\rho=\frac12\left(P^{11}_{11}+P^{22}_{11}+P^{12}_{11}+(P^{12}_{11})^\dagger\right)
\end{equation}

Now, we must consider that, as Rob is causally disconnected from Ridler's region $IV$, we should trace over that region to obtain Rob's density matrix. Hence,  we need to compute the trace over $IV$ for each of the previous operators \eqref{0000},\eqref{1111},\eqref{0011}.

Taking this trace is actually quite straightforward taking into account the orthonormality of our basis once we have chosen one particular ordering criterion \eqref{ordering}, 
\begin{equation}\label{products}
\braket{\tilde m}{\tilde m'}_{IV}=\delta_{mm'}\left(\delta_{s_1,s'_1}\delta_{k_1,k'_1}\dots\delta_{s_m,s'_m}\delta_{k_m,k'_m}\right) 
\end{equation} 
Hence,
\begin{equation}\label{traza00}
\tr_{IV} P_{00}=\sum_{m'=0}^{2n}\bra{\tilde m'}_{IV}P_{00}\ket{\tilde m'}_{IV}.
\end{equation}
Using \eqref{products} only the diagonal elements in region $IV$ survive and \eqref{traza00} turns out to be
\begin{equation}\label{t00pre}
\tr_{IV} P_{00}=\sum_{m=0}^{2n}|C^m|^2 \!\!\!\sum_{\substack{s_1,\dots,s_m\\k_1,\dots,k_m}}\!\!\!\!\xi_{s_1,\dots,s_m}^{k_1,\dots,k_m}\ket{0}_A\ket{\tilde m}_{I}\langle{\tilde m}|_{I}\bra{0}_A
\end{equation}
which, substituting $C^m$ as a function of $C^0$ using \eqref{coeff2} and then \eqref{Des}, is expressed as
\begin{equation}\label{t00}
\tr_{IV} P_{00}=\sum_{m=0}^{2n}D_0^m \!\!\!\sum_{\substack{s_1,\dots,s_m\\k_1,\dots,k_m}}\!\!\!\!\xi_{s_1,\dots,s_m}^{k_1,\dots,k_m}\ket{0}_A\ket{\tilde m}_{I}\langle{\tilde m}|_{I}\bra{0}_A
\end{equation}

Now we will compute the trace 
\begin{equation}\label{traza11}
\tr_{IV} P^{ij}_{11}=\sum_{m'=0}^{2n}\bra{\tilde m'}_{IV}P^{ij}_{11}\ket{\tilde m'}_{IV}
\end{equation}
\begin{eqnarray}\label{t11pre}
\nonumber \tr_{IV} P_{11}^{ij}&=&\sum_{m=0}^{2n-1}|A^m|^2 \!\!\!\sum_{\substack{s_1,\dots,s_m\\k_1,\dots,k_m}}\!\!\!\!\xi_{s_1,\dots,s_m,s^i_R}^{k_1,\dots,k_m,k^i_R}\xi_{s_1,\dots,s_m,s^j_R}^{k_1,\dots,k_m,k^j_R}\\*
&&\!\!\!\!\!\!\times\ket{k_A,s_A}_A\ket{\tilde m;k_R,s_R}_{I}\langle{\tilde m;k_R,s_R}|_{I}\bra{k_A,s_A}_A\nonumber\\
\end{eqnarray}
substituting $C^m$ as a function of $C^0$ (combining \eqref{Am} and \eqref{coeff2}) we can express $|A^m|^2$ as
\begin{equation}
|C_0|^2 \tan^{2m}r\left(\cos r+\frac{\sin^2 r}{\cos r}\right)^2=|C_0|^2 \frac{\tan^{2m}r}{\cos^2 r}=D^m_2
\end{equation}
Such that
\begin{eqnarray}\label{t11}
\nonumber \tr_{IV} P^{ii}_{11}&=&\sum_{m=0}^{2n-1}D^m_2 \!\!\!\sum_{\substack{s_1,\dots,s_m\\k_1,\dots,k_m}}\!\!\!\!\xi_{s_1,\dots,s_m,s_R}^{k_1,\dots,k_m,k_R}\ket{k^i_A,s^i_A}_A\\*
&&\!\!\!\!\!\!\!\!\times\ket{\tilde m;k^i_R,s^i_R}_{I}\langle{\tilde m;k^j_R,s^j_R}|_{I}\bra{k^j_A,s^j_A}_A
\end{eqnarray}
When $\{k^i_R,s^i_R\}=\{k^j_R,s^j_R\}$ $\equiv$ $\{k_R,s_R\}$, $\{k^i_A,s^i_A\}=\{k^j_A,s^j_A\}\equiv \{k_A,s_A\}$.

However, in the general case $i\neq j$ it would be
\bibliographystyle{apsrev}
\begin{eqnarray}\label{t112}
\nonumber \tr_{IV} P^{ij}_{11}&=&\sum_{m=0}^{2n-1}D^m_2 \!\!\!\sum_{\substack{s_1,\dots,s_m\\k_1,\dots,k_m}}\!\!\!\!\xi_{s_1,\dots,s_m,s^i_R,s^j_R}^{k_1,\dots,k_m,k^i_R,k^j_R}\ket{k_A,s_A}_A\\*
&&\!\!\!\!\!\!\!\!\times\ket{\tilde m;k_R,s_R}_{I}\langle{\tilde m;k_R,s_R}|_{I}\bra{k_A,s_A}_A
\end{eqnarray}

Now, let us compute the trace
\begin{equation}\label{traza01pre}
\tr_{IV} P_{01}=\sum_{m'=0}^{2n}\bra{\tilde m'}_{IV}P_{01}\ket{\tilde m'}_{IV}
\end{equation}
\begin{eqnarray}\label{traza01pre2}
\nonumber \tr_{IV} P_{01}\!\!&=&\!\!\sum_{m=0}^{2n-1}C^m (A^{m})^*\!\!\!\sum_{\substack{s_1,\dots,s_{m}\\k_1,\dots,k_{m}}}\!\!\!\!\xi_{s_1,\dots,s_m,s_R}^{k_1,\dots,k_m,k_R}\ket{0}_A\biket{\tilde m}{\tilde m}\\*
\!\!&& \!\!\times\langle{\tilde l}|_{IV}\langle{\tilde l;k_R,s_R}|_{I}\bra{0}_A
\end{eqnarray}
from \eqref{Am} and \eqref{coeff2} we see that the product $C^m(A^m)^*$ is real and has the expression $C^m(A^m)^*=$
\begin{equation}
|C_0|^2\tan^{2m}r\left(\cos r+\frac{\sin^2 r}{\cos r}\right)=|C_0|^2\frac{\tan^{2m}r}{\cos r}=D^m_1
\end{equation}
so that
\begin{eqnarray}\label{traza01}
\nonumber \tr_{IV} P_{01}\!\!&=&\!\!\sum_{m=0}^{2n-1}D^m_1\!\!\!\sum_{\substack{s_1,\dots,s_{m}\\k_1,\dots,k_{m}}}\!\!\!\!\xi_{s_1,\dots,s_m,s_R}^{k_1,\dots,k_m,k_R}\ket{0}_A\biket{\tilde m}{\tilde m}\\*
\!\!&& \!\!\times\langle{\tilde l}|_{IV}\langle{\tilde l;k_R,s_R}|_{I}\bra{0}_A
\end{eqnarray}

Now we can compute Rob's density matrices for each case tracing over IV in expressions \eqref{roimp1} and \eqref{roimp2}. First the matrix \eqref{roimp1} is, after tracing over IV, 
\begin{equation}
\tr_{IV}\rho=\frac12\left(\tr_{IV}P_{00}+\tr_{IV}P_{01}+\tr_{IV}P_{01}^\dagger+\tr_{IV}P^{ii}_{11}\right)
\end{equation}
substituting expressions \eqref{t00}, \eqref{t112}, \eqref{traza01} we get expression \eqref{densmat1}.

Now, concerning \eqref{roimp2}
\begin{equation}
\tr_{IV} \rho=\frac12\tr_{IV}\left( P^{11}_{11}+P^{22}_{11}+P^{12}_{11}+(P^{12}_{11})^\dagger\right)
\end{equation}
Substituting expressions \eqref{t11} and \eqref{t112} we obtain expression \eqref{densmat2}.

The derivation of \eqref{densmat3} is completely analogous to \eqref{densmat1}, taking now into account that we have $\hat C^m$ and $\hat D^m$ instead of $C^m$ and $D^m$ and that we have no spin degree of freedom. Notice that, even though the structure of \eqref{densmat3} is completely analogous to the structure of \eqref{densmat1}, and therefore, repeating the derivation will add nothing to this appendix,  these density matrices are completely different due to the different dimensions, the different values of $\hat C^0$ and $C^0$ and the number of $2\times 2$ blocks which give negative eigenvalues. 
\bibliographystyle{apsrev}

\end{document}